\documentclass[conference,a4paper]{IEEEtran}

\usepackage[utf8]{inputenc} 
\usepackage[T1]{fontenc}
\usepackage{url, balance}
\usepackage{ifthen}
\usepackage[cmex10]{amsmath} 

\usepackage{amssymb} 
\usepackage{mathtools} 
\usepackage{amsfonts} 
\usepackage{accents} 

\usepackage{mleftright}\mleftright 

\usepackage{multirow}

\usepackage[dvips]{graphicx} 
\graphicspath{{./}}
\DeclareGraphicsExtensions{.eps}

\usepackage[a4paper,margin=0.75in,footskip=20pt]{geometry} 

\usepackage[dvipsnames,table]{xcolor} 

\usepackage{bbm} 
\usepackage{xfrac} 
\usepackage{cancel} 
\usepackage{wrapfig} 
\usepackage{tensor} 

\usepackage{theorem}
\usepackage[backend=bibtex,bibstyle=ieee,citestyle=numeric-comp,url=false,doi=false,isbn=false]{biblatex}
\usepackage{url} 

\usepackage{psfrag} 

\usepackage{tikz}
\usetikzlibrary{positioning}

\usepackage[inline]{enumitem} 

\usepackage{array} 

\usepackage{aliascnt} 

\usepackage[draft]{hyperref} 

\usepackage{cleveref} 

\usepackage{orcidlink} 

\usepackage[ruled,algoruled,vlined]{algorithm2e}

\SetCommentSty{mycommfont}

\SetKwInput{KwInput}{Input}                
\SetKwInput{KwOutput}{Output}              
\SetKwRepeat{Do}{do}{while}%
\SetKwProg{Init}{init}{}{}

\usepackage{xparse} 

\usepackage{ifthen} 

\usepackage{comment}

\crefname{section}{Section}{Sections}
\crefname{subsection}{Section}{Sections}
\crefname{equation}{Eq.}{Equations}
\crefname{enumi}{part}{parts}
\crefname{table}{Table}{Tables}
\crefname{figure}{Figure}{Figures}
\crefname{algocf}{Algorithm}{Algorithms}

\newtheorem{theorem}{Theorem}
\crefname{theorem}{Theorem}{Theorems}

\newaliascnt{lemma}{theorem}
\newtheorem{lemma}[lemma]{Lemma}
\aliascntresetthe{lemma}
\crefname{lemma}{Lemma}{Lemmas}

\newaliascnt{definition}{theorem}
\newtheorem{definition}[definition]{Definition}
\aliascntresetthe{definition}
\crefname{definition}{Definition}{Definitions}

\newaliascnt{corollary}{theorem}

\aliascntresetthe{corollary}
\crefname{corollary}{Corollary}{Corollarys}

\newaliascnt{claim}{theorem}

\aliascntresetthe{claim}
\crefname{claim}{Claim}{Claims}

\newaliascnt{conjecture}{theorem}

\aliascntresetthe{conjecture}
\crefname{conjecture}{Conjecture}{Conjectures}

\newaliascnt{question}{theorem}
\newtheorem{question}[question]{Question}
\aliascntresetthe{question}
\crefname{question}{Question}{Questions}

\newaliascnt{problem}{theorem}

\aliascntresetthe{problem}
\crefname{problem}{Problem}{Problems}

\newaliascnt{example}{theorem}
\newtheorem{example}[example]{Example}
\aliascntresetthe{example}
\crefname{example}{Example}{Examples}

\newaliascnt{oquestion}{theorem}

\aliascntresetthe{oquestion}
\crefname{oquestion}{Open Question}{Open Questions}

\theoremstyle{plain}
\theorembodyfont{\normalfont}

\newaliascnt{remark}{theorem}
\newtheorem{remark}[remark]{Remark}
\aliascntresetthe{remark}
\crefname{remark}{Remark}{Remark}

\newtheorem{cnstr}{Construction}

\newenvironment{construction}{\begin{cnstr}}{\hfill$\Box$\end{cnstr}}
\crefname{cnstr}{Construction}{Constructions}

\crefname{step}{Step}{Steps}

\crefname{regime}{Regime}{Regimes}

\newtheorem{myalgo}{Algorithm}

\crefname{myalgo}{Algorithm}{Algorithms}

\newcounter{enumrom}
\renewcommand{\theenumrom}{(\roman{enumrom})}

\makeatletter
\renewcommand{\@endtheorem}{\endtrivlist}
\makeatother

\makeatletter
\renewcommand{\thefigure}{{\@arabic\c@figure}}
\renewcommand{\fnum@figure}{{\bf Figure\,\thefigure}}
\makeatother

\renewcommand{\leq}{\leqslant}
\renewcommand{\geq}{\geqslant}

\newcommand{\bfa}{{\boldsymbol a}}
\newcommand{\bfb}{{\boldsymbol b}}

\newcommand{\bfu}{{\boldsymbol u}}
\newcommand{\bfv}{{\boldsymbol v}}
\newcommand{\bfw}{{\boldsymbol w}}
\newcommand{\bfx}{{\boldsymbol x}}
\newcommand{\bfy}{{\boldsymbol y}}
\newcommand{\bfz}{{\boldsymbol z}}

\newcommand{\cC}{\mathcal{C}}

\newcommand{\cG}{\mathcal{G}}

\newcommand{\cI}{\mathcal{I}}

\newcommand{\cQ}{\mathcal{Q}}

\newcommand{\Ftwo}{{{\Sigma}}_{\!2}}

\newcommand{\bfal}{{\boldsymbol \alpha}}
\newcommand{\bfbe}{{\boldsymbol \beta}}

\DeclarePairedDelimiter\abs{\lvert}{\rvert}

\DeclarePairedDelimiter\parenv{\lparen}{\rparen}
\DeclarePairedDelimiter\sparenv{\lbrack}{\rbrack}
\DeclarePairedDelimiter\bracenv{\lbrace}{\rbrace}

\DeclarePairedDelimiterX\mathset[2]{\lbrace}{\rbrace}{#1 : #2}
\DeclarePairedDelimiterX\multset[2]{\lbrace\!\!\lbrace}{\rbrace\!\!\rbrace}{#1 : #2}
\DeclarePairedDelimiterX\inner[2]{\langle}{\rangle}{#1 \mathrel{},\mathrel{} #2}
\DeclarePairedDelimiterX\condparenv[2]{(}{)}{#1 \mathrel{}\delimsize\vert\mathrel{} #2}

\DeclareDocumentCommand\norm{ o m }{
\IfNoValueTF{#1}
{\left\Vert#2\right\Vert}
{\left\Vert#2\right\Vert_{#1}}
}

\DeclareDocumentCommand\der{ o m o }{
\IfNoValueTF{#1}
{
\IfNoValueTF{#3}
{\frac{d}{d{#2}}}
{\frac{d{#3}}{d{#2}}}
}
{\parenv*{\frac{d}{d{#2}}}^{#1}\IfNoValueTF{#3}{}{#3}}
}
\DeclareDocumentCommand\partder{ o m m }{
\IfNoValueTF{#1}
{\frac{\partial{#3}}{\partial{#2}}}
{\frac{\partial^{#1}{#3}}{{\partial{#2}}^{#1}}}
}
\DeclareDocumentCommand\df{ o m o }{
d\IfNoValueTF{#1}{}{^{#1}}{#2}\IfNoValueTF{#3}{}{_{#3}}
}

\DeclareMathOperator{\tr}{tr}

\DeclareMathOperator{\wt}{wt}

\DeclarePairedDelimiter{\ceil}{\lceil}{\rceil}
\DeclarePairedDelimiter{\floor}{\lfloor}{\rfloor}

\DeclareDocumentCommand\enc{ o }{
\IfNoValueTF{#1}
{\operatorname{Enc}}
{\operatorname{Enc}_{\ref*{#1}}}
}

\DeclareDocumentCommand\dec{ o }{
\IfNoValueTF{#1}
{\operatorname{Dec}}
{\operatorname{Dec}_{\ref*{#1}}}
}

\newtheorem{defn}[theorem]{Definition}

\makeatletter
\newcommand\code[1]{%
\@ifundefined{r@#1}{%
\cC_{\operatorname*{#1}}%
}{%
\cC_{\ref*{#1}}%
}%
}
\makeatother

\bibliography{literature.bib}

\renewcommand{\tr}[2][]{\mathcal{R}_{#1}(#2)}
\newcommand{\tri}[3][]{\tr[#1]{\boldsymbol{#2}}_{#3}}

\usepackage{fancyhdr}

\IEEEoverridecommandlockouts
\begin{document}
\title{Correcting Multiple Substitutions in Nanopore-Sequencing Reads

\thanks{
  The research was Funded by the European Union (ERC, DNAStorage, 101045114 and EIC, DiDAX 101115134). Views and opinions expressed are however those of the authors only and do not necessarily reflect those of the European Union or the European Research Council Executive Agency. Neither the European Union nor the granting authority can be held responsible for them.  
  It was also partially funded by UKRI BBSRC grant BB/Y007638/1, the Department for Science, Innovation and Technology (DSIT) and the Royal Academy of Engineering for a Chair in Emerging Technologies award.}%
}


\author{%
  \IEEEauthorblockN{Anisha Banerjee\IEEEauthorrefmark{1},
                    Yonatan Yehezkeally\IEEEauthorrefmark{2},
                    Antonia Wachter-Zeh\IEEEauthorrefmark{1},
                    and Eitan Yaakobi\IEEEauthorrefmark{3}}
  \IEEEauthorblockA{\IEEEauthorrefmark{1}%
                    Institute for Communications Engineering, Technical University of Munich (TUM), Munich, Germany}
  \IEEEauthorblockA{\IEEEauthorrefmark{2}%
                    School of Computing, 
                    Newcastle University, 
                    Newcastle upon Tyne NE4 5TG, United Kingdom}
  \IEEEauthorblockA{\IEEEauthorrefmark{3}%
                    Department of Computer Science, Technion---Israel  Institute  of  Technology, Haifa 3200003, Israel}
    \IEEEauthorblockA{Email: anisha.banerjee@tum.de, 
      yonatan.yehezkeally@ncl.ac.uk, 
      antonia.wachter-zeh@tum.de,
      yaakobi@cs.technion.ac.il}
      \\[-5.8ex]

}


\maketitle

\pagestyle{empty}
\thispagestyle{fancy}


\begin{abstract}
    Despite 
    their significant advantages over competing technologies, nanopore sequencers are plagued by high error rates
    , due to 
    physical characteristics of the nanopore and 
    inherent noise in the biological processes
    . It is thus paramount not only to formulate efficient error-correcting constructions for these channels, but also to establish bounds on the minimum redundancy required by such coding schemes. In this context, we adopt a simplified model of nanopore sequencing inspired by the work of Mao \emph{et al.}, 
    accounting for 
    the effects of intersymbol interference and measurement noise. For a
    n input sequence of length $n$, 
    The 
    vector that is produced, designated as the \emph{read vector}, may additionally suffer at most \(t\) substitution errors. We employ the well-known graph-theoretic clique-cover technique to establish that 
    at least \(t\log n -O(1)\) bits of redundancy are required to correct multiple (\(t \geq 2\)) substitutions. While this is surprising in comparison to the case of a single 
    substitution, that necessitates at most \(\log \log n - O(1)\) bits of redundancy, a suitable error-correcting code that is optimal up to a constant follows immediately from the properties of 
    read vectors. 
\end{abstract}

\section{Introduction}

DNA as a potential data storage medium holds great promise. However, significant advancements in synthesis and sequencing technologies are still necessary to make it feasible for commercial use. Among the various sequencing technologies, nanopore sequencing outshines its contenders due to its 
portability and ability to support longer reads. This technology sequences a DNA strand by allowing it to pass through a microscopic pore that contains $\ell$ nucleotides at any given time instant. By analyzing the variations in the ionic current, which are influenced by the different nucleotides passing through the pore, we can infer the sequence of nucleotides in the original DNA strand. This 
readout, however, is plagued by 
distortions due to the noise inherent in the different physical aspects of this process. To begin with, because the pore can hold multiple nucleotides (\( \ell > 1 \)) simultaneously, the observed current is influenced by several nucleotides rather than just one
creating 
inter-symbol interference (ISI) in the channel output. Additionally, the irregular movement of the DNA fragment through the pore can often lead to backtracking or to skipping a few nucleotides. These irregularities appear as duplications or deletions in the channel output, respectively. Furthermore, the random noise that affects the measured current can be effectively modeled as substitution errors.

Initial work in this area focused on devising accurate mathematical models for the sequencer or formulating efficient error-correcting codes that incorporate these models. Notably, \cite{maoModels2018, hulettCoding2021, mcbainFiniteState2022a, mcbainInformationRatesNoisy2024} examined the channel from an information-theoretic perspective in an effort to understand its capacity. In particular, the authors of \cite{maoModels2018} proposed a channel model that considers ISI, deletions, and random measurement noise and also derived suitable upper bounds for the capacity of this channel. On the contrary, \cite{hulettCoding2021} used a more deterministic model and developed an algorithm to calculate its capacity. The authors of \cite{mcbainFiniteState2022a, mcbainInformationRatesNoisy2024} studied a finite-state semi-Markov channel (FSMC)--based model for nanopore sequencing that accounts for ISI, duplications, and noisy measurements. They estimated the achievable information rates of this noisy nanopore channel by formulating efficient algorithms to perform this computation. Another exciting direction seeks to facilitate the accurate decoding of DNA fragments, despite sample duplications and background noise, by designing codes directly based on the current signals produced by the nanopore for each sequence of nucleotides \cite{vidalConcatenatedNanoporeDNA2024, vidalErrorBoundsDecoding2023, vidalUnionBoundGeneralized2023}.

A subset of prior work \cite{banerjeeCorrectingSingleDeletion2024a, banerjeeErrorCorrectingCodesNanopore2024, cheeCodingSchemeNoisy2024, sunBoundsConstructionsEll2024} studied a specific channel model of the nanopore sequencer inspired by \cite{maoModels2018}. In particular, these works considered the nanopore sequencer to be the concatenation of three distinct channels, as depicted in Fig. 1. The first component that emulates the ISI effect is parameterized by $\ell$ and demonstrates how the observed current is influenced by the $\ell>1$ consecutive nucleotides present in the pore at any moment. This stage is conceptualized as a window of length $\ell$ that slides over an input sequence, shifting by a single position after each time step. This produces a sequence of $\ell$-mers, or a string of $\ell$ symbols, which is then fed to a discrete memoryless channel (DMC) that transforms each $\ell$-mer into a discrete voltage level based on a deterministic function, assumed to be the Hamming weight when the input is binary, and the composition function otherwise. The final stage introduces errors, such as substitutions, deletions, or duplications, that corrupt the sequence of discrete voltage levels. While \cite{banerjeeCorrectingSingleDeletion2024a, banerjeeErrorCorrectingCodesNanopore2024} assumed that the error component introduces either at most one substitution or at most one deletion in the final channel output, \cite{sunBoundsConstructionsEll2024} considered multiple substitutions and investigated bounds and constructions of suitable substitution-correcting codes for this channel, when \(\ell = 2\). In contrast, \cite{cheeCodingSchemeNoisy2024} proposed constrained codes for this channel to combat duplication and deletion errors. This model is also similar to the transverse-read channel \cite{cheeTransverseReadCodesDomainWall2023, yerushalmiCapacityWeightedRead2024}, which is important for racetrack memories. We now state the problem formally. 
\begin{question}
    For a given input \( \bfx \), let \( \tr[\ell]{\bfx} \) denote the error-free output from the channel (as defined in Definition~\ref{def::read-vec}). What is the minimum redundancy required to correct \( t \) substitution errors in \( \tr[\ell]{\bfx} \) (rather than in \( \bfx \))?
\end{question}

The primary contribution of this work is to demonstrate that for \( \ell \geq 2 \), the minimum redundancy required for any code of length $n$ that can correct \( t \geq 2 \) substitutions in \( \ell \)-read vectors is at least \( t\log_2 n -O(1) \) bits (Theorem~\ref{th::t-sub-red}), which starkly contrasts with the minimum redundancy bound of \( \log_2\log_2 n - o(1) \) \cite{banerjeeErrorCorrectingCodesNanopore2024} for the case of $t=1$. Following the introduction of essential notations and definitions in Section~\ref{sec::prelim}, we present the proof of this lower bound, based on the clique cover technique \cite{knuthSandwich1994, chrisnataCorrecting2022}, in Section~\ref{sec::mult_sub}. With this unfortunate result, we show in \Cref{constr::t-sub-code} that a naive construction achieves this bound up to a constant.


\begin{figure}[t]
	\scalebox{0.75}{
	\begin{tikzpicture}
	
	\node[align=center] (start) {Nucleotides \\ $\bfx$};
	
	\node[draw, minimum width=1.3cm,
	minimum height=2.6cm, right=0.6cm of start, rounded corners, align=center] (ISI) {ISI \\ $\ell$ };
	
	\node[draw, minimum width=1.3cm,
	minimum height=2.6cm, right=14mm of ISI, rounded corners] (dmc) {DMC};
	
	\node[draw, minimum width=1.4cm,
	minimum height=2.6cm, right=16mm of dmc, rounded corners] (del) {Substitution};
	
	\node[align=center, right=0.6cm of del] (end) {};

	\draw[->] (start) -- (ISI);
	\draw[->] (ISI) --  node[above=0.3pt]{$\ell$-mers} (dmc);
    \draw[->] (dmc) -- node[above, align=center]{Discrete \\ voltage \\ levels} node[below=3.0pt]{$\tr{\bfx}$} (del);
 
	\draw[->] (del) -- (end) node[align=center, right=0.1pt] (h) {$\tr{\bfx}'$};
	
	\end{tikzpicture}}
	\caption{Simplified model of a nanopore sequencer }
    
	\label{fig::ch_model}
\end{figure}
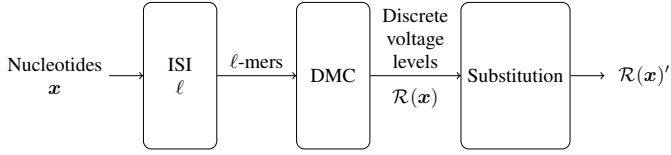

\section{Preliminaries} \label{sec::prelim}

For any $q \geq 2$, we let $\Sigma_q$ represent the $q$-ary alphabet $\{0,1,\ldots,q-1\}$. The set of all $q$-ary sequences of length $n$ is indicated by $\Sigma_q^n$, with $\Sigma_q^0$ meant to indicate the empty set.  Additionally, we let $\Sigma_q^{\geq n} = \cup_{i=n}^{\infty} \Sigma_q^i$. For any two integers $a,b$ such that $a\leq b$, $[a,b]$ is used to denote the set $\{a,a+1,\ldots, b\}$, and $[b] \triangleq [1,b]$ for $b \geq 1$. We also use the following notation to denote element-wise modulo operation on a vector $\bfy \in \Sigma_q^n$.
\begin{align*}
\bfy \bmod a\triangleq\big( y_1\bmod a, y_2\bmod a, \ldots, y_n\bmod a \big).
\end{align*}

Given any vector $\bfx=(x_1, \ldots, x_n)$, its substring $(x_i, x_{i+1}, \ldots, x_j)$ is indicated concisely as $\bfx_i^j$. The operator $\wt(\bfx)$ indicates the Hamming weight of $\bfx$, while $|\bfx|$ refers to its length, i.e., $|\bfx|=n$. We also indicate the Hamming distance between any two vectors $\bfx, \bfy \in \Sigma_q^n$ as 
\begin{equation*}
d_H(\bfx, \bfy)=|\{i : i\in [n], x_i \neq y_i\}|.
\end{equation*}

Throughout this work, we focus on ${q=2}$, and the output of the channel we consider is a sequence of readings of a sliding window moved across $\bfx$, as defined below.
\begin{defn} \cite{banerjeeCorrectingSingleDeletion2024a, banerjeeErrorCorrectingCodesNanopore2023} \label{def::read-vec}
    The $\ell$-\emph{read vector} of any ${\bfx \in \Ftwo^n}$ is of length $n+\ell-1$ over $\Sigma_{\ell+1}$, and is defined as
\begin{equation*}
\tr[\ell]{\bfx} \triangleq (\wt(\bfx_{2-\ell}^1), \wt(\bfx_{3-\ell}^{2}), \ldots, \wt(\bfx_{n}^{n+\ell-1})), \label{def::read-vec}
\end{equation*}
where for any $i\not\in [n]$, we set $x_i=0$. 
\end{defn}

The $i$-th element of $\tr[\ell]{\bfx}$ is denoted as $\tr[\ell]{\bfx}_i$; that is, $\tri[\ell]{x}{i}=\wt(\bfx_{i-\ell+1}^{i})$. We will omit $\ell$ from the subscript of $\tr[\ell]{\bfx}$ whenever it is clear from the context. It is worth pointing out that Definition~\ref{def::read-vec} can be extended to the non-binary alphabet, by considering compositions instead of Hamming weights \cite{banerjeeErrorCorrectingCodesNanopore2024}. The composition of a vector $q$-ary \( \bfx \) refers to the count of each symbol in $\Sigma_q$ as it appears in \(\bfx\).

\begin{example}
    The $3$-read vector of $\bfx=(0,1,1,0,1,0)$ is $\tr{\bfx}=(0,1,2,2,2,1,1,0)$. Its fourth element is $\tri{\bfx}{4} = 2$. \label{eg::read}
\end{example}

As mentioned previously, \cite{cheeTransverseReadCodesDomainWall2023, yerushalmiCapacityWeightedRead2024} examine a similar model, wherein the output sequence, called the transverse-read vector, is essentially a substring of the $\ell$-read vector for specific parameter choices.


Since we are interested in codes that correct up to $t$ substitutions in $\ell$-read vectors, it is essential to precisely define what is meant by an error-correcting code in the framework of our channel. Similarly to \cite{banerjeeErrorCorrectingCodesNanopore2024}, a code is said to be a \emph{$t$-substitution $\ell$-read code} if for any distinct $\bfx, \bfy$ from this code, it holds that $d_H(\tr{\bfx}, \tr{\bfy})>2t$.

\section{Correcting Multiple Substitutions} \label{sec::mult_sub}

This section aims to establish an upper bound on the size of a code that corrects~$t$ substitutions in $\ell$-read vectors, where $t \geq 2$ is constant; the case $t=1$ was thoroughly analysed in \cite{banerjeeErrorCorrectingCodesNanopore2024}. To accomplish this, we apply the clique cover technique, which was also used in \cite{chrisnataCorrecting2022, banerjeeErrorCorrectingCodesNanopore2024} for the case of $t = 1$
. This method considers a graph \( \cG(n) \) that contains vertices corresponding to all vectors in \( \Sigma^n_2 \). In this graph, any two vertices representing distinct binary vectors, say $\bfx$ and $\bfy$ in \( \Sigma^n_2 \), are considered adjacent if and only they satisfy \( d_H(\tr{\bfx}, \tr{\bfy}) \leq 2t\). Consequently, any subset of vertices in \( \cG(n) \) where no two vertices are adjacent (namely, an \emph{independent set}) constitutes a $t$-substitution $\ell$-read code. In contrast, a \emph{clique} in the graph is a set of vertices that are all pair-wise adjacent. The formal definition of a clique \emph{cover} is stated below.
\begin{definition}
	A \textbf{clique cover} $\mathcal{Q}$ is a collection of cliques in a graph $\mathcal{G}$, such that every vertex in $\mathcal{G}$ belongs to at least one clique in $\mathcal{Q}$.
\end{definition}

From~\cite{chrisnataOptimal2020a, knuthSandwich1994}, the following result is widely known.
\begin{theorem}
	If $\mathcal{Q}$ is a clique cover in a graph~$\mathcal{G}$, then the size of any independent set is at most $|\mathcal{Q}|$. \label{th::cq_cover}
\end{theorem}

This theorem implies 
that the size of a clique cover is also an upper bound on the cardinality of a $t$-substitution $\ell$-read code. Hence, we seek to define an appropriate clique cover $\cQ$ for the remainder of this section. To proceed along these lines, we first define the following permutation, which 
serves to simplify the presentation of the technical results laid out in Lemma~\ref{lem::gamma_i_clique} and Lemma~\ref{lem::dh-2t}.


\begin{definition}\label{defn:pi_p}
    \cite[Definition~3]{banerjeeErrorCorrectingCodesNanopore2023}
    For a positive integer~$p$, define a permutation~$\pi_p$ on $\Ftwo^n$ as follows. For all $\bfx\in \Ftwo^n$, arrange the coordinates of $\bfx_1^{p\ell \floor{n/(p\ell)}}$ in a matrix $X\in \Sigma^{p\floor{n/(p\ell)}\times \ell}$, by row (first fill the first row from left to right, then the next, etc.). 
    Next, partition $X$ into sub-matrices of dimension $p\times 2$ (if $\ell$ is odd, we ignore $X$'s right-most column). 
    Finally, going through each sub-matrix (from left to right, and then top to bottom), we concatenate its \emph{rows}, to obtain $\pi_p(\bfx)$ (where unused coordinates from~$\bfx$ are appended arbitrarily).
    
    More precisely, for all $0\leq i < \floor{\frac{n}{p \ell}}$, $0\leq j < \floor{\frac{\ell}{2}}$ and $0\leq k < p$ denote 
    \[
    \bfx^{(i,j,k)} =\linebreak x_{(i p+k)\ell+2j+1} x_{(i p+k)\ell+2j+2};
    \]
    then 
    \[
    \bfx^{(i,j)} = \bfx^{(i,j,0)}\circ \cdots\circ \bfx^{(i,j,p-1)}
    \]
    and 
    \[
    \bfx^{(i)} = \bfx^{(i,0)}\circ \cdots\circ \bfx^{(i,\floor{\ell/2}-1)}.
    \]
    Then $\pi_p(\bfx) = \bfx^{(0)}\circ \cdots\circ \bfx^{(\floor{n/p \ell}-1)} \circ \tilde{\bfx}$, where $\tilde{\bfx}$ is composed of all coordinates of $\bfx$ not earlier included.

    Let the function $f_{\pi} : [n] \rightarrow [n]$ map a coordinate of $\pi_p(\bfx)$ onto $\bfx$, i.e., for all $i \in [n]$, we have $\pi_p(\bfx)_i = \bfx_{f_{\pi}(i)}$. 
\end{definition}


\begin{remark} \label{rem::sum-x-r}
    Consider some $\bfx \in \Sigma_2^n$ and a positive integer $p$ for which $\pi_p(\bfx)=\bfu \; \circ \; \bfal^m \; \circ \; \bfv$, where $\bfu,\bfv \in \Sigma_2^{\geq 0}$, $|\bfu| \equiv 0 \pmod{2p}$, $m \geq 1$ and $\bfal \in  \{01,10\}$. Thus, $\bfx$ has the form
    \begin{IEEEeqnarray*}{+rCl+x*}
        \bfx &=& \bfu' \; \circ \; \bfal \; \circ \; \bfw_1 \; \circ \; \bfal \;\circ \; \cdots \; \bfw_{m-1} \; \circ \; \bfal \; \circ \; \bfv',
    \end{IEEEeqnarray*}
    where $\bfu'$ has even length and for all $h \in [m-1]$, $\bfw_h \in \Sigma_2^{\ell-2}$. Let $c = |\bfu| + 1$ denote the index at which the substring $\bfal^m$ starts in $\pi_p(\bfx)$ and for any $i \in  [m-1]$, let $r = f_{\pi}(c+2i-1) = |\bfu'| + (i-1)\ell+2$. Observe that
    \begin{IEEEeqnarray*}{+rCl+x*}
        x_{r} + x_{r+\ell-1} &=& x_{|\bfu'| + (i-1)\ell+2} + x_{|\bfu'| + i\ell+1} \\
        &=& \pi_p(\bfx)_{|\bfu|+2i}+ \pi_p(\bfx)_{|\bfu|+2i+1}\\
        &=& \alpha_2 + \alpha_1 = 1.
    \end{IEEEeqnarray*}
\end{remark}

\begin{example} \label{eg::def-pi}
	    Consider $\bfx=(0,1,1,0,1,0)$ and ${\bfy=(1,0,1,1,0,0)}$. For $p=2$ and $\ell=3$,
    \begin{IEEEeqnarray*}{+rCl+x*}
		X&=\begin{bmatrix}
		0 & 1 & 1 \\
		0 & 1 & 0
		\end{bmatrix}\; =& \begin{bmatrix}
		\bfx^{(0,0,0)} & x_3 \\
		\bfx^{(0,0,1)} & x_6 
		\end{bmatrix}, \\        
        Y&=\begin{bmatrix}
		1 & 0 & 1 \\
		1 & 0 & 0
		\end{bmatrix} =& \begin{bmatrix}
		\bfy^{(0,0,0)} & y_3 \\
		\bfy^{(0,0,1)} & y_6
		\end{bmatrix}.
	\end{IEEEeqnarray*}
	Since $\ell$ is odd, the last columns of $X$ and $Y$ are ignored. Upon partitioning the respective results into $2 \times 2$ sub-matrices, we get $\pi_p(\bfx)=(1,0,1,0,1,0)$ and $\pi_p(\bfy)=(0,1,0,1,1,0)$ (unused coordinates were appended based on the order of their indices). One can see that $f_{\pi}(1)=1$, $f_{\pi}(2)=2$, $f_{\pi}(3)=4$, $f_{\pi}(4)=5$, $f_{\pi}(5)=3$ and $f_{\pi}(6)=6$. 

    Since $\pi_p(\bfx)=(01)^2 \circ 10$ and $\pi_p(\bfy)=(10)^2 \circ 10$, we note in the context of Remark~\ref{rem::sum-x-r}, that for $r=f_{\pi}(2)=2$, it holds that $x_r+x_{r+\ell-1} = y_r+y_{r+\ell-1}=1$.
\end{example}

The subsequent definition presents the core component of our clique cover, and borrows ideas from \cite{chrisnataCorrecting2022, banerjeeErrorCorrectingCodesNanopore2024}. 
\begin{definition} \label{def::cq}
	For a positive integer~$p$, let 
    \begin{align*}
    \Lambda^{(1)}_p &= \mathset*{(01)^{j} (10)^{p-j}}{j\in [p]}, \\
    \Lambda^{(2)}_p &= \mathset*{(10)^{j} (01)^{p-j}}{j\in [p]} \\
    \Lambda_p &= \Lambda^{(1)}_p \cup \Lambda^{(2)}_p,
    \end{align*}
	where $\bfa^0=\bfb^0$ is the empty word, and $\widetilde{\Lambda}_p = \Ftwo^{2p}\setminus \Lambda_p$. Further, let $m = \floor{\frac{\ell}{2}} \floor{\frac{n}{p\ell}}$ and
    \begin{IEEEeqnarray*}{+rCl+x*}
        \Gamma_1 &=& \mathset*{(\bfu, \bfw)}{i \in [m], \bfu\in \widetilde{\Lambda}_p^{i-1}, \bfw\in \widetilde{\Lambda}_p^{m-i}},   \\
        \Gamma_2 &=& \{(\bfu, \bfv, \bfw) : i_1, i_2 \in [m], i_1+i_2 \leq m, \\
        && \quad \bfu\in \widetilde{\Lambda}_p^{i_1-1},  \bfv\in \widetilde{\Lambda}_p^{i_2-1}, \bfw\in \widetilde{\Lambda}_p^{m-i_1-i_2} \}, \\
        \vdots && \qquad \qquad \qquad \qquad  \vdots \\
        \Gamma_k &=& \{(\bfu, \bfv_1, \ldots, \bfv_{k-1}, \bfw) : h \in [k], i_h \in [m], \sum_{r=1}^{k} i_r \leq m, \\  && \quad \bfu \in \widetilde{\Lambda}_p^{i_1-1}, \bfv_h \in \widetilde{\Lambda}_p^{i_{h+1}-1}, \bfw\in \widetilde{\Lambda}_p^{m-\sum_{r=1}^{k} i_r} \}, \\
        \vdots && \qquad \qquad \qquad \qquad  \vdots \\
        \Gamma_t &=& \{(\bfu, \bfv_1, \ldots, \bfv_{t-1}, \bfw) : h \in [t], i_h \in [m], \sum_{r=1}^{t} i_r \leq m, \\
        && \quad \bfu \in \widetilde{\Lambda}_p^{i_1-1}, \bfv_h \in \widetilde{\Lambda}_p^{i_{h+1}-1}, \bfw\in \Ftwo^{2p(m-\sum_{r=1}^{t} i_r)} \},
    \end{IEEEeqnarray*}
    where $\widetilde{\Lambda}_p^0$ is the singleton that contains an empty word. Then, for all $i \in [t]$ and all $\gamma = (\bfu, \bfv_1, \ldots, \bfv_{i-1}, \bfw) \in \Gamma_i$, define    \begin{IEEEeqnarray*}{+rCl+x*}
       Q_{\gamma} &=& \{ \bfu (\bfal_1)^{h_1} (\bfbe_1)^{p - h_1} \bfv_1 \cdots \bfv_{i-1} (\bfal_i)^{h_i} (\bfbe_i)^{p - h_i} \bfw \\
       && \quad : h_1, \ldots, h_i \in [p]\},
    \end{IEEEeqnarray*}
    where for all $r \in [i]$, $\{\bfal_r, \bfbe_r\} = \{01, 10\}$. There are clearly $2^i$ such distinct sets, each corresponding to a specific choice of the tuple $(\bfal_1, \ldots, \bfal_i) \in \{01,10\}^i$. We index these sets as $Q_{\gamma}^{(0)},\ldots, Q_{\gamma}^{(2^i - 1)}$ and let 
	\begin{IEEEeqnarray*}{+rCl+x*}
        \mathcal{Q}(m,p) &= \mathset*{\bracenv*{\bfx}}{\bfx\in \widetilde{\Lambda}_p^m} \cup \mathset*{Q_\gamma^{(0)},Q_\gamma^{(1)}}{\gamma\in \Gamma_1}
	\cup \cdots \\
    & \quad  \cdots \cup\mathset*{Q_\gamma^{(0)},\cdots, Q_\gamma^{(2^t - 1)}}{\gamma\in \Gamma_t}.
    \end{IEEEeqnarray*}	
\end{definition}

In what follows, we endeavor to show that $\cQ(m,p)$ maps onto a clique cover of $\cG(2pm)$, as a stepping stone to presenting the clique cover for the larger graph $\cG(n)$.

\begin{lemma} \label{lem::gamma_i_clique}
    Consider two distinct vectors $\bfx, \bfy \in \Ftwo^{2pm}$ such that there exist integers $p \geq 1$ and $s\in [2,t]$ for which $\pi_{p}(\bfx)$ and $\pi_{p}(\bfy)$ are related as follows.
    \begin{IEEEeqnarray*}{+rCl+x*}
        \pi_p(\bfx) &= \bfu \circ \bfa_1 \circ \bfv_1\circ \cdots \circ \bfv_{s-1} \circ \bfa_s \circ \bfw \\
        \pi_p(\bfy) &= \bfu \circ \bfb_1  \circ \bfv_1\circ \cdots \circ \bfv_{s-1} \circ \bfb_s  \circ \bfw,    
    \end{IEEEeqnarray*}
    where for all $i \in [s - 1]$, $\bfv_i \in (\widetilde{\Lambda}_p)^{\geq 0}$, $\bfu \in (\widetilde{\Lambda}_p)^{\geq 0}$ $\bfw \in \Ftwo^{\geq 0}$ where $m = \floor{\frac{\ell}{2}} \floor{\frac{2m}{\ell}}$, for all $j \in [s]$, either $\bfa_j, \bfb_j \in \Lambda^{(1)}_p$ or $\bfa_j, \bfb_j \in \Lambda^{(2)}_p$; and $\bfa_j \neq \bfb_j$. Then, it holds that $d_H(\tr{\bfx}, \tr{\bfy}) \leq 2s$. 
\end{lemma}

\begin{IEEEproof}
It follows from the definitions of $\Lambda_p^{(1)}$ and $\Lambda_p^{(2)}$ that there exist some $\bfu',\bfv'_1, \ldots, \bfv'_{s-1}, \bfw' \in \Ftwo^{\geq 0}$ and integers $m_1, \ldots, m_s \in [1,p-1]$ such that $\pi_p(\bfx)$ and $\pi_p(\bfy)$ are also expressible in the following form
\begin{IEEEeqnarray*}{+rCl+x*}
	\pi_p(\bfx) &= \bfu' \circ (\bfal_1)^{m_1} \circ \bfv'_1 \circ \cdots \circ \bfv'_{t-1} \circ (\bfal_s)^{m_s} \circ \bfw', \\
	\pi_p(\bfy) &= \bfu' \circ (\bfbe_1)^{m_1}  \circ \bfv'_1\circ \cdots \circ \bfv'_{t-1} \circ (\bfbe_s)^{m_s}  \circ \bfw',    
\end{IEEEeqnarray*}
where for all $h \in [s]$, $\{\bfal_h, \bfbe_h\} = \{01, 10\}$. 

Observe that we either have $\bfu' = \bfu \; \circ \; (01)^{r}$ 
or $\bfu' = \bfu \; \circ \; (10)^{r}$ (depending on whether $\{\bfal_1, \bfbe_1\} \in \Lambda_p^{(1)}$ or $\{\bfal_1, \bfbe_1\} \in \Lambda_p^{(2)}$) where $r \geq 1$. Since $\bfu \in  (\widetilde{\Lambda}_p)^{\geq 0}$, it is easy to see that $|\bfu'| \equiv 0 \pmod 2$. This in combination with Definition~\ref{defn:pi_p} implies that for all $(i,j,k) \in [0, \floor{\frac{2m}{\ell}} - 1] \times [0, \floor{\frac{\ell}{2}} - 1] \times [0, p-1]$, either $\bfx^{(i,j,k)} = \bfy^{(i,j,k)}$ or $\{ \bfx^{(i,j,k)}, \bfy^{(i,j,k)} \} = \{01, 10\}$. In both cases, it holds that $\wt(\bfx^{(i,j,k)}) = \wt(\bfy^{(i,j,k)})$. 

We begin by proving the lemma statement for even values of $\ell$. Notably, when $r \leq 2pm - \ell +1$ and $r$ is odd, there exist integers $i_1, \ldots, i_{\ell / 2}, j_1, \ldots, j_{\ell / 2}, k_1, \ldots, k_{\ell / 2}$ that satisfy
\begin{IEEEeqnarray*}{+rCl+x*}
	\wt(\bfx_{r}^{r+\ell-1}) &=& \wt(\bfx^{(i_1,j_1,k_1)}) + \cdots + \wt(\bfx^{(i_{\ell / 2},j_{\ell / 2},k_{\ell / 2})}) \\
	&=& \wt(\bfy^{(i_1,j_1,k_1)}) + \cdots + \wt(\bfy^{(i_{\ell / 2},j_{\ell / 2},k_{\ell / 2})}) \\
	&=& \wt(\bfy_{r}^{r+\ell-1}),
\end{IEEEeqnarray*}
i.e., $\tri{\bfx}{r+\ell-1} = \tri{\bfy}{r+\ell-1}$. This also holds when $r>2pm - \ell +1$ and $r$ is odd, since $x_i=y_i=0$ for all $i\not\in[2pm]$. On the contrary, when $r$ is even, similar arguments lead us to
\begin{IEEEeqnarray}{+rCl+x}
	\tri{\bfx}{r+\ell-1}\! - \!\tri{\bfy}{r+\ell-1} \! &=& \! \wt(\bfx_{r}^{r+\ell-1}) - \wt(\bfy_{r}^{r+\ell-1}) \nonumber \\
    &=& \! x_{r+\ell-1} - y_{r+\ell-1} \! - x_{r} + y_{r}. \label{eq::trd}
\end{IEEEeqnarray}

To specify the set of indices in $[2pm+\ell-1]$ at which $\tr{\bfx}$ and $\tr{\bfy}$ disagree, we let the index at which the substring $(\bfal_h)^{m_h}$ (or $(\bfbe_h)^{m_h}$) starts in $\pi_p(\bfx)$ (or $\pi_p(\bfy)$) be given by $c_h$, for all $h\in [s]$. Thus, the starting and ending indices of each $(\bfal_h)^{m_h}$ (or $(\bfbe_h)^{m_h}$) in $\pi_p(\bfx)$ (or $\pi_p(\bfy)$) are $c_h$ and $c_h+2m_h-1$, which map to the positions $f_{\pi}(c_h)$ and $f_{\pi}(c_h+2m_h-1)$ in $\bfx$ (or $\bfy$), respectively. Since for all $i \in \bigcup_{h \in [s]} \{f_{\pi}(c_h), \ldots, f_{\pi}(c_h+2m_h-1)\}$, $x_i \neq y_i$, it is possible that for certain instances of $r$, $\tri{\bfx}{r+\ell-1} \neq \tri{\bfy}{r+\ell-1}$.

Observe from Remark~\ref{rem::sum-x-r}, that for any $r \in \bigcup_{h \in [s]} \{f_{\pi}(c_h+1),f_{\pi}(c_h+3), \ldots, f_{\pi}(c_h+2m_h-3)\}$ (even integers when $\ell$ is even and equivalent to $r+\ell-1 \in \bigcup_{h \in [s]} \{f_{\pi}(c_h+2), f_{\pi}(c_h+4), \ldots \}$), we have $x_r+x_{r+\ell-1} =y_r+y_{r+\ell-1} = 1$. Thus, the only interesting cases that remain, are when either $r \in \bigcup_{h \in [s]} \{f_{\pi}(c_h+2m_h-1)\}$ or when $r+\ell-1 \in \bigcup_{h \in [s]} \{f_{\pi}(c_h)\}$. In other words, we have $\tri{\bfx}{r+\ell-1} \neq \tri{\bfy}{r+\ell-1}$ only if $r \in  \bigcup_{h \in [s]} \{f_{\pi}(c_h+2m_h-1), f_{\pi}(c_h) -\ell+1\}$, which is a set of size $2s$. Consequently, $d_H(\tr{\bfx}, \tr{\bfy}) \leq 2s$. 

To prove the same for odd values of $\ell$, we infer from Definition~\ref{defn:pi_p} 
that for all $r \in [2pm]$ satisfying $r \equiv 0 \pmod \ell$, we have $x_r = y_r$. We continue as before, by noting that for any $r \leq (\floor{2m/\ell}p-1)\ell+1$ that satisfies $r \bmod \ell \in \{0,1,3,\ldots, \ell-2\}$, there exist integers $i_1, \ldots, i_{\floor{\ell/2}}, j_1, \ldots, j_{\floor{\ell/2}}, k_1, \ldots, k_{\floor{\ell/2}}$, that satisfy
\begin{IEEEeqnarray*}{+rCl+x*}
	&&\tri{\bfx}{r+\ell-1} = \wt(\bfx_r^{r+\ell-1})\\
    &=&x_{\ceil{r/\ell} \ell} +
    \wt(\bfx^{(i_1,j_1,k_1)}) + \cdots + \wt(\bfx^{(i_{\floor{{\ell}/{2}}},j_{\floor{\ell/2}},k_{\floor{\ell/2}})}) \\
	&=& y_{\ceil{r/\ell} \ell} +
    \wt(\bfy^{(i_1,j_1,k_1)}) + \cdots  + \wt(\bfy^{(i_{\floor{{\ell}/{2}}},j_{\floor{\ell/2}},k_{\floor{\ell/2}})})\\
	&=& \tri{\bfy}{r+\ell-1}.
\end{IEEEeqnarray*}

The same holds when $r>(\floor{2m/\ell}p-1)\ell+1$ and $r \bmod \ell \in \{0,1,3,\ldots, \ell-2\}$ as $x_i=y_i=0$ for all $i\not\in[2pm]$, and similarly so $r\geq \floor{2m/\ell}p\ell+1$. The remaining case to examine is when $r \bmod \ell \in \{2,4,\ldots, \ell-1\}$, we deduce upon applying similar arguments that (\ref{eq::trd}) holds also for odd values of $\ell$, and ultimately conclude similarly from Remark~\ref{rem::sum-x-r} that $d_H(\tr{\bfx}, \tr{\bfy}) \leq 2s$. 
\end{IEEEproof}
\begin{example}
For $\ell =3$, $p=2$ and the vectors $\bfx=(0,1,1,0,1,0)$ and ${\bfy=(1,0,1,1,0,0)}$ (from Example~\ref{eg::def-pi}), the substring $(01)^2$ (or $(10)^2$) starts in $\pi_p(\bfx)$ (or $\pi_p(\bfy)$) at index $c=1$. Observe that $s=1$ and for $r \in \{f_{\pi}(c)-\ell+1, f_{\pi}(c+3)\}=\{-1, 5\}$, we have $\tri{\bfx}{r+\ell-1}\neq\tri{\bfy}{r+\ell-1}$, i.e., $d_H(\tr{\bfx}, \tr{\bfy})=2s=2$.
\end{example}

With the assistance of Lemma~\ref{lem::gamma_i_clique}, we are now ready to show that $\mathcal{Q}(m,p)$ is a clique cover of the smaller graph $\cG(2p m)$, where $m = \floor{\frac{\ell}{2}} \floor{\frac{n}{p\ell}}$.
\begin{lemma} \label{lem::dh-2t}
	Letting $m = \floor{\frac{\ell}{2}} \floor{\frac{n}{p\ell}}$, 
    \begin{align*}
        \mathset*{\pi_p^{-1}(Q)}{Q\in \mathcal{Q}(m,p)}
    \end{align*}
    is a clique cover of $\cG(2p m)$.
\end{lemma}
(Here, we abuse notation to let $\pi_p$ also act on $\Ftwo^{2pm}$, in the natural way.)
\begin{IEEEproof}
	While the singletons forming the set $\mathset*{\bracenv*{\bfx}}{\pi_p(\bfx)\in \widetilde{\Lambda}_p^m}$ are evidently cliques, Lemma~\ref{lem::gamma_i_clique} implies that for all $i \in [2,t], \gamma\in \Gamma_i$, and $0\leq j<2^i$, 
    \begin{align*}
        \mathset*{\bfx\in \Ftwo^{2pm}}{\pi_p(\bfx)\in Q_\gamma^{(j)}}
    \end{align*}
    is also a clique (the case $i=1$ was already proven in \cite{banerjeeErrorCorrectingCodesNanopore2023}). It now remains to show that $\pi_p(\bfx)$ belongs to at least one clique in $\mathcal{Q}(m,p)$ for any $\bfx \in \Ftwo^{2pm}$. For simplicity, we use the fact that $\pi_p$ is a permutation to show instead that any such $\bfx$ is itself a member.
	
	To this end, consider a $\bfx \in \Ftwo^{2pm}$ and the set $\cI = \{i_1, \ldots, i_{|\cI|}\}$ where $i_1 < \cdots < i_{|\cI|}$, that satisfies $i_h - 1 \equiv 0 \pmod{2p}$ and  $\pi_p(\bfx)_{i_h}^{i_h+2p-1} \in \Lambda_p$ for all $h\in [|\cI|]$. We consider $\cI$ to be exhaustive, i.e., there exists no $i \in [2p(m -1) + 1]$ such that $i - 1 \equiv 0 \pmod{2p}$, $\pi_p(\bfx)_{i}^{i+2p-1} \in \Lambda_p$ and $i\not\in\cI$.

    If $\cI$ is empty, $\bfx$ forms a singleton. If however $0 < |\cI| < t$, then $\bfx$ belongs to a clique in $\Gamma_{|\cI|}$, say $Q_{\gamma}$ for some $\gamma=(\bfu, \bfv_1, \ldots, \bfv_{|\cI| -1}, \bfw)$, such that $\bfu = \pi_p(\bfx)_1^{i_1-1}$, $\bfw = \pi_p(\bfx)_{i_{|\cI|}+2p}^{2pm}$ and for all $h \in [|\cI|-1]$, $\bfv_h = \pi_p(\bfx)_{i_h+2p}^{i_{h+1}-1}$. When $t \leq |\cI| \leq m$, $\bfx$ belongs to a clique in $\Gamma_{t}$, say $Q_{\gamma'}$ for some $\gamma'=(\bfu', \bfv'_1, \ldots, \bfv'_{t -1}, \bfw')$, where $\bfu = \pi_p(\bfx)_1^{i_1-1}$, $\bfw = \pi_p(\bfx)_{i_{t}+2p}^{2pm}$ and for all $h \in [t-1]$, $\bfv_h = \pi_p(\bfx)_{i_h+2p}^{i_{h+1}-1}$. Thus, each $\bfx \in \Ftwo^{2pm}$ belongs to at least one clique in $\mathcal{Q}(m,p)$.
\end{IEEEproof}
Our next step is to adapt the clique cover $\cQ(m,p)$ over the smaller graph $\cG(2pm)$, to construct a clique cover for $\cG(n)$.
\begin{theorem} \label{th::t-sub-cq-cover}
	Let 
	\[
	\mathcal{Q}_p = \mathset*{\pi_p^{-1}(Q\times \bracenv*{\bfz})}{Q\in \mathcal{Q}(m,p), \bfz\in \Ftwo^{n-2p m}}, 
	\]
	where $\pi_p^{-1}(A) = \mathset*{\bfu\in \Ftwo^n}{\pi_p(\bfu)\in A}$. Then, $\mathcal{Q}_p$ is a clique cover in $\mathcal{G}(n)$.
\end{theorem}
\begin{IEEEproof}
It readily follows from $\bigcup \mathcal{Q}(m,p) = \Sigma_2^{2p m}$ that $\bigcup \mathcal{Q}_p = \Sigma_2^n$. Lemma~\ref{lem::dh-2t} proves that every element of $\mathcal{Q}_p$ is a clique of $\mathcal{G}(n)$. This concludes the proof.
\end{IEEEproof}
Recall that we have constructed the clique cover $\cQ_p$ in order to bound the minimum redundancy of any $t$-substitution $\ell$-read code
. We therefore proceed to compute its size.
\begin{lemma} \label{lem::num-cliques}
    The total number of cliques is given by
    \begin{IEEEeqnarray*}{+rCl+x*}
    \abs*{\mathcal{Q}_p}
    &=& 2^{n} \sparenv*{ \sum_{i=0}^{t-1} 2^i \binom{m}{i} \frac{\lambda^{m-i}}{2^{2pi}} + 2^{-(2p-1)t} \sum_{r=0}^{m-t} \binom{r+t-1}{t-1} \lambda^r } 
    \end{IEEEeqnarray*}
	where $m = \floor{\frac{\ell}{2}} \floor{\frac{n}{p\ell}}$ and $\lambda = 1-\frac{2p}{2^{2p}} $. 
\end{lemma}
\begin{IEEEproof}
    From Definition~\ref{def::cq}, it follows that the number of singletons equals $|\widetilde{\Lambda}_p|^m$ where $|\widetilde{\Lambda}_p| = 2^{2p} - 2p$. 
    
    Also recall from Definition~\ref{def::cq} that for a particular $\gamma \in \Gamma_i$, there exist $2^i$ distinct cliques $Q_{\gamma}^{(0)}, \ldots, Q_{\gamma}^{(2^i-1)}$. Thus, the number of cliques (excluding singletons) is given by $\sum_{i=1}^{t} 2^i |\Gamma_i|$ 
    , where for $i \in [t - 1]$, $|\Gamma_i| = \binom{m}{i} |\widetilde{\Lambda}_p|^{m-i}$ and 
    \begin{IEEEeqnarray*}{+rCl+x*}
       |\Gamma_t| &=& \sum_{i_1, \ldots, i_t} |\widetilde{\Lambda}|^{i_1+\cdots+i_t-t} 2^{2p(m-i_1 - \cdots - i_t)} \\
       &=& \sum_{r=t}^{m} \binom{r-1}{t-1} |\widetilde{\Lambda}_p|^{r-t} 2^{2p(m-r)} \\
       &=& 2^{2pm} \sum_{r=t}^{m} \binom{r-1}{t-1} |\widetilde{\Lambda}_p|^{r-t} 2^{-2pr} \\
       &=& 2^{2p(m-t)} \sum_{r=0}^{m-t} \binom{r+t-1}{t-1} |\widetilde{\Lambda}_p|^{r} 2^{-2pr}.
    \end{IEEEeqnarray*}

    We let $\lambda = \frac{|\widetilde{\Lambda}_p|}{2^{2p}} = (1-\frac{2p}{2^{2p}})$. This leads to
    \begin{IEEEeqnarray*}{+rCl+x*}
	   \abs*{\mathcal{Q}(m,p)} &=& \sum_{i=0}^{t-1} 2^i \binom{m}{i} |\widetilde{\Lambda}_p|^{m-i} \\
       && + 2^{t + 2p(m-t)} \sum_{r=0}^{m-t} \binom{r+t-1}{t-1} |\widetilde{\Lambda}_p|^{r} 2^{-2pr} \\
       &=& 2^{2pm} \Big[ \sum_{i=0}^{t-1} 2^i \binom{m}{i} \frac{\lambda^{m-i}}{2^{2pi}} \\
       && + 2^{-(2p-1)t} \sum_{r=0}^{m-t} \binom{r+t-1}{t-1} \lambda^r \Big]. 
    \end{IEEEeqnarray*}

    The previous equation, coupled with the fact that $|\cQ_p| = 2^{n-2pm}|\cQ(m,p)|$ leads to the statement of the lemma.    
\end{IEEEproof}

It follows from \Cref{lem::num-cliques} that
\begin{IEEEeqnarray}{+rCl+x*}
    \log_2 \abs*{\mathcal{Q}_p} &=& n - (2p-1) t + \log_2 \sparenv*{ \sum_{r=0}^{m-t} \binom{r+t-1}{t-1} \lambda^r } \nonumber\\
    && + \log_2 \sparenv*{ 1 + \frac{ \sum_{i=0}^{t-1} 2^i \binom{m}{i} \frac{\lambda^{m-i}}{2^{2pi}} }{ \sum_{r=0}^{m-t} \binom{r+t-1}{t-1} \lambda^r  }  }. \label{eq::log-qp}
\end{IEEEeqnarray}

We simplify and bound the latter components below.

\begin{lemma} \label{lem::quant1}
    For $s \geq 0$, $p = \ceil{\frac{1}{2}(1-\epsilon) \log_2 n}$ and $ 0 < \lambda < 1$, 
    \begin{IEEEeqnarray*}{+rCl+x*}
    \lim_{n \to \infty} \sum_{i=0}^{s} \binom{m}{i} \frac{\lambda^{m-i}}{2^{(2p-1)i}} &=&0,
    \end{IEEEeqnarray*}
    where $m = \floor{\frac{\ell}{2}} \floor{\frac{n}{p\ell}}$.
\end{lemma}
\begin{IEEEproof} 
Observe that since $2^{2pi} \geq n^{i(1-\epsilon)}$ and $m \leq n / 2p$,
\begin{IEEEeqnarray*}{+rCl+x*}
    \binom{m}{i} \frac{\lambda^{m-i}}{2^{(2p-1)i}} 
    &\leq& \binom{m}{i} \frac{\lambda^{m-i}}{n^{i(1-\epsilon)}2^{-i}} \\
    & < & \frac{m^i}{i!}\frac{\lambda^{m-i}}{n^{i(1-\epsilon)}2^{-i}} \leq \frac{1}{i!} \frac{\lambda^{m-i}}{n^{-i\epsilon} p^i} \\
    &\leq& \frac{1}{i!} \frac{2^i}{(1-\epsilon)^i} \frac{\lambda^{m-i}}{n^{-i\epsilon} (\log_2 n)^i},
\end{IEEEeqnarray*}
where the final inequality follows from $p \geq \frac{1}{2}(1-\epsilon) \log_2 n$. 
Since the decay rate of $\lambda^{m-i}$ dominates, we infer that $\lim_{n \to \infty} \frac{\lambda^{m-i}}{n^{-i\epsilon} (\log_2 n)^i} = 0$ and the lemma follows.
\end{IEEEproof}

\begin{lemma} \label{lem::quant2}
    For $0 < \lambda< 1$ and a positive integer $t$, it holds that 
    \begin{IEEEeqnarray*}{+rCl+x*}
    \lim_{n \to \infty} \sum_{r=0}^{n} \binom{r+t-1}{t-1} \lambda^r &=& O(1).
    \end{IEEEeqnarray*}
\end{lemma}
\begin{IEEEproof}
    We apply the following inequality \cite{bar-levAdversarialTornPaperCodes2023} 
    \begin{IEEEeqnarray*}{+rCl+x*}
    \log_2 \binom{u+v}{u} &\leq& u (2\log (e) + \log (\frac{v}{u}) ),
    \end{IEEEeqnarray*}
    to deduce that $\binom{r+t-1}{t-1} \lambda^r \leq  \big(\frac{e^2}{t-1}\big)^{t-1} r^{t-1} \lambda^r$.
    Note that $r^{t-1} \lambda^r = r^{t-1} e^{r\log \lambda}$ is maximized when $(t-1)r^{t-2} \lambda^r + r^{t-1}\lambda^r \log \lambda = 0$, i.e., $r= - (t-1) / \log \lambda$. Similarly, $r^{t-1} \lambda^{r/2}$ achieves its maximum when $r= -2 (t-1) / \log \lambda$. This allows us to bound the following summation.
    \begin{IEEEeqnarray*}{+rCl+x*}
        \IEEEeqnarraymulticol{3}{l}{\lim_{n \to \infty} \sum_{r=0}^{n} r^{t-1} \lambda^r} \\
        &\leq& \frac{t-1}{\log (1/\lambda)} \parenv*{\frac{t-1}{\log (1/\lambda)} \lambda^{-1/\log \lambda}}^{t-1} \\
        && + \lim_{n \to \infty} \sum_{r = 1 + \frac{t-1}{\log(1/\lambda)}}^{n} \parenv*{\frac{2(t-1)}{\log (1/\lambda)} \lambda^{-1/\log \lambda}}^{t-1} \lambda^{r/2} \\
         &=& \frac{t-1}{\log (1/\lambda)}  + \parenv*{\frac{2(t-1)}{e\log (1/\lambda)}}^{t-1}  \frac{\lambda^{{1/2-(t-1)/(2\log \lambda})}}{1-\lambda^{1/2}} \\
         &=& \parenv*{\frac{t-1}{e\log (1/\lambda)}}^{t-1} \parenv*{\frac{t-1}{\log (1/\lambda)} + 2^{t-1}\frac{\lambda^{1/2} e^{-(t-1)/2} }{1-\lambda^{1/2}} }. 
    \end{IEEEeqnarray*}
    This 
   implies that $\lim_{n \to \infty} \sum_{r=0}^{n} \binom{r+t-1}{t-1} \lambda^r$ is also finite.
\end{IEEEproof}

The application of Lemma~\ref{lem::quant1} and Lemma~\ref{lem::quant2} to (\ref{eq::log-qp}) finally yields the following bound on the minimum redundancy of any $t$-substitution $\ell$-read code.
\begin{theorem} \label{th::t-sub-red}
    The redundancy of any $t$-substitution $\ell$-read code, for $t,\ell\geq 2$, is bounded from below by
    \[
        t\log_2 n - O(1).
    \]
\end{theorem}

This theorem suggests that for $\ell \geq 2$ and $t\geq 2$, the minimum redundancy required by any $t$-substitution-correcting code for a channel that produces $\ell$-read vectors is, up to a fixed addend, the same as that of the classical substitution channel \cite[Theorem~4.3, Lemma~4.8]{rothIntroductionCodingTheory2007}%
, assuming $t$ is fixed 
with respect to $n$
. For $\ell=2$, this result is also proved in \cite{sunBoundsConstructionsEll2024}. Theorem~\ref{th::t-sub-red} is dispiriting in light of 
\cite[Lemma~14]{banerjeeErrorCorrectingCodesNanopore2024}, which shows that to correct a single ($t=1$) substitution in $\ell$-read vectors, when $\ell \geq 3$, $\log_2\log_2 n + o(1)$ redundant bits are sufficient. 

Next, we focus on designing a suitable error-correcting code by leveraging the fact that $\bfx$ can be directly inferred from the first or last $n$ elements of $\tr{\bfx} \bmod 2$ \cite[Proposition~1]{banerjeeErrorCorrectingCodesNanopore2023}. This leads us to the following naive $t$-substitution $\ell$-read code construction is optimal up to a constant.
\begin{construction} \label{constr::t-sub-code}
    \begin{IEEEeqnarray*}{+rCl+x*}
    \{ \bfx \in \Sigma_2^n :  (\tri{\bfx}{1}, \ldots, \tri{\bfx}{n}) \bmod 2 \in \cC(n,t)
    \},
\end{IEEEeqnarray*}
where $\cC(n,t) \subset \Sigma_2^n$ is a $t$-substitution-correcting code, i.e., for any distinct $\bfx,\bfy \in \cC(n,t)$, it holds that $d_H(\bfx,\bfy) > 2t$.    
\end{construction}
Evidently, \Cref{constr::t-sub-code} requires $t\log_2 n$ redundant bits. Thus, for $t\geq 2$ and $\ell \geq 2$, it is a $t$-substitution $\ell$-read code that is optimal up to a constant.

\section{Conclusion}

This work uses a simplified model of a nanopore sequencer and establishes a lower bound on the redundancy needed to correct up to $t$ substitutions in the output of this simplified channel. Our findings indicate that for \(t \geq 2\), the minimal redundancy for a suitable code is comparable to that of a classical substitution channel. This prompts the question of whether these results would still hold if the channel model assigned non-uniform weights to the bits in each window, i.e., $\tri{\bfx}{i} = \sum_{h=1}^{\ell} w_h x_{i-\ell+h}$ and $\bfw \neq (1,\ldots,1)$. 

\balance
\printbibliography

@book{rothIntroductionCodingTheory2007,
  title = {Introduction to Coding Theory},
  author = {Roth, Ron M.},
  year = {2007},
  publisher = {Cambridge University Press},
  isbn = {978-0-521-84504-5}
}

@article{mcbainInformationRatesNoisy2024,
  title = {Information Rates of the Noisy Nanopore Channel},
  author = {McBain, Brendon and Viterbo, Emanuele and Saunderson, James},
  year = {2024},
  month = aug,
  journal = {IEEE Transactions on Information Theory},
  volume = {70},
  number = {8},
  pages = {5640--5652},
  issn = {1557-9654},
  doi = {10.1109/TIT.2024.3404002},
  urldate = {2025-01-14}
}

@misc{sunBoundsConstructionsEll2024,
  title = {Bounds and Constructions of \${\textbackslash}ell\$-Read Codes under the Hamming Metric},
  author = {Sun, Yubo and Ge, Gennian},
  year = {2024},
  month = mar,
  number = {arXiv:2403.11754},
  eprint = {2403.11754},
  primaryclass = {cs, math},
  publisher = {arXiv},
  doi = {10.48550/arXiv.2403.11754},
  urldate = {2024-04-02},
  archiveprefix = {arXiv}
}

@article{bar-levAdversarialTornPaperCodes2023,
  title = {Adversarial Torn-Paper Codes},
  author = {{Bar-Lev}, Daniella and Marcovich, Sagi and Yaakobi, Eitan and Yehezkeally, Yonatan},
  year = {2023},
  month = oct,
  journal = {IEEE Transactions on Information Theory},
  volume = {69},
  number = {10},
  pages = {6414--6427},
  issn = {0018-9448, 1557-9654},
  doi = {10.1109/TIT.2023.3292895},
  urldate = {2025-01-20},
  copyright = {https://ieeexplore.ieee.org/Xplorehelp/downloads/license-information/IEEE.html}
}

@inproceedings{cheeCodingSchemeNoisy2024,
  title = {Coding Scheme for Noisy Nanopore Sequencing with Backtracking and Skipping Errors},
  booktitle = {2024 IEEE International Symposium on Information Theory (ISIT)},
  author = {Chee, Yeow Meng and Immink, Kees A. Schouhamer and Vu, Van Khu},
  year = {2024},
  month = jul,
  pages = {458--463},
  issn = {2157-8117},
  doi = {10.1109/ISIT57864.2024.10619277},
  urldate = {2025-01-14}
}

@inproceedings{banerjeeCorrectingSingleDeletion2024a,
  title = {Correcting a Single Deletion in Reads from a Nanopore Sequencer},
  booktitle = {2024 IEEE International Symposium on Information Theory (ISIT)},
  author = {Banerjee, Anisha and Yehezkeally, Yonatan and {Wachter-Zeh}, Antonia and Yaakobi, Eitan},
  year = {2024},
  month = jul,
  pages = {103--108},
  issn = {2157-8117},
  doi = {10.1109/ISIT57864.2024.10619468},
  urldate = {2024-09-18},
  copyright = {All rights reserved}
}

@inproceedings{banerjeeErrorCorrectingCodesNanopore2023,
  title = {Error-Correcting Codes for Nanopore Sequencing},
  booktitle = {{{IEEE Intl. Symp.}} {{Inf. Theory}} ({{ISIT}})},
  author = {Banerjee, Anisha and Yehezkeally, Yonatan and {Wachter-Zeh}, Antonia and Yaakobi, Eitan},
  year = {2023},
  month = jun,
  pages = {364--369},
  publisher = {{IEEE}},
  address = {{Taipei, Taiwan}},
  doi = {10.1109/ISIT54713.2023.10206710},
  isbn = {978-1-66547-554-9}
}

@article{banerjeeErrorCorrectingCodesNanopore2024,
  title = {Error-Correcting Codes for Nanopore Sequencing},
  author = {Banerjee, Anisha and Yehezkeally, Yonatan and {Wachter-Zeh}, Antonia and Yaakobi, Eitan},
  year = {2024},
  month = jul,
  journal = {IEEE Tran. Inf. Theory},
  volume = {70},
  number = {7},
  pages = {4956--4967},
  issn = {1557-9654},
  doi = {10.1109/TIT.2024.3380615},
  urldate = {2024-04-02},
  copyright = {All rights reserved}
}

@article{cheeTransverseReadCodesDomainWall2023,
  title = {Transverse-Read-Codes for Domain Wall Memories},
  author = {Chee, Yeow Meng and Vardy, Alexander and Vu, Van Khu and Yaakobi, Eitan},
  year = {2023},
  journal = {IEEE Journal on Selected Areas in Inf. Theory},
  volume = {4},
  pages = {784--793},
  issn = {2641-8770},
  doi = {10.1109/JSAIT.2023.3334303},
  urldate = {2024-04-16}
}

@inproceedings{chrisnataOptimal2020a,
	address = {Kapolei, HI, USA},
	title = {Optimal Reconstruction Codes for Deletion Channels},
	author = {Chrisnata, Johan and Kiah, Han Mao and Yaakobi, Eitan},
	month = oct,
	pages = {279--283},
    year = {2020},
    booktitle = {{IEEE} Intl. Symp. Inf. Theory Appl. (ISITA)},
}

@article{knuthSandwich1994,
	title = {The Sandwich Theorem},
	volume = {1},
	issn = {1077-8926},
	doi = {10.37236/1193},
	number = {1},
	urldate = {2022-08-12},
	journal = {The Electronic Journal of Combinatorics},
	author = {Knuth, Donald E.},
	month = apr,
	year = {1994},
	pages = {A1},
}

@article{chrisnataCorrecting2022,
	title = {Correcting Deletions With Multiple Reads},
	volume = {68},
	issn = {0018-9448, 1557-9654},
	doi = {10.1109/TIT.2022.3184868},
	number = {11},
	urldate = {2022-11-17},
	journal = {IEEE Trans. Inf. Theory},
	author = {Chrisnata, Johan and Kiah, Han Mao and Yaakobi, Eitan},
	month = nov,
	year = {2022},
	pages = {7141--7158},
}

@article{maoModels2018,
	title = {Models and Information-Theoretic Bounds for Nanopore Sequencing},
	volume = {64},
	issn = {0018-9448, 1557-9654},
	doi = {10.1109/TIT.2018.2809001},
	number = {4},
	urldate = {2022-03-09},
	journal = {IEEE Trans. Inf. Theory},
	author = {Mao, Wei and Diggavi, Suhas N. and Kannan, Sreeram},
	month = apr,
	year = {2018},
	pages = {3216--3236},
}

@inproceedings{hulettCoding2021,
	address = {Melbourne, Australia},
	title = {On coding for an Abstracted Nanopore Channel for {DNA} Storage},
	isbn = {978-1-5386-8209-8},
	doi = {10.1109/ISIT45174.2021.9518236},
	urldate = {2022-12-08},
	booktitle = {{IEEE} {Intl.} {Symp.} on {Inf.} {Theory} ({ISIT})},
	author = {Hulett, Reyna and Chandak, Shubham and Wootters, Mary},
	month = jul,
	year = {2021},
	pages = {2465--2470},
}

@inproceedings{mcbainFiniteState2022a,
	address = {Espoo, Finland},
	title = {Finite-State Semi-Markov Channels for Nanopore Sequencing},
	isbn = {978-1-66542-159-1},
	doi = {10.1109/ISIT50566.2022.9834633},
	urldate = {2022-12-08},
	booktitle = {{IEEE} {Intl.} {Symp.} {Inf.} {Theory} ({ISIT})},
	author = {McBain, Brendon and Viterbo, Emanuele and Saunderson, James},
	month = jun,
	year = {2022},
	pages = {216--221},
}

@article{vidalConcatenatedNanoporeDNA2024,
  title = {Concatenated {{Nanopore DNA Codes}}},
  author = {Vidal, Adrian and Wijekoon, V. B. and Viterbo, Emanuele},
  year = {2024},
  month = apr,
  journal = {IEEE Tran. on NanoBioscience},
  volume = {23},
  number = {2},
  pages = {310--318},
  issn = {1558-2639},
  doi = {10.1109/TNB.2024.3350001},
  urldate = {2024-04-16}
}

@inproceedings{yerushalmiCapacityWeightedRead2024,
  address = {Accepted Apr 2024},
  title = {The Capacity of the Weighted Read Channel},
  author = {Yerushalmi, Omer and Etzion, Tuvi and Yaakobi, Eitan},
  pages = {(arXiv preprint arXiv:2401.15368)},
  booktitle = {Proc. {IEEE} {Intl.} {Symp.} {Inf.} {Theory} ({ISIT})}
}

@inproceedings{vidalUnionBoundGeneralized2023,
  title = {Union Bound for Generalized Duplication Channels with DTW Decoding},
  booktitle = {2023 IEEE Intl. Symp. Inf. Theory (ISIT)},
  author = {Vidal, Adrian and Wijekoon, V. B. and Viterbo, Emanuele},
  year = {2023},
  month = jun,
  pages = {358--363},
  issn = {2157-8117},
  doi = {10.1109/ISIT54713.2023.10206475},
  urldate = {2024-03-14}
}

@inproceedings{vidalErrorBoundsDecoding2023,
  title = {Error Bounds for Decoding Piecewise Constant Nanopore Signals in DNA Storage},
  booktitle = {ICC 2023 - IEEE Intl. Conf. Comm.},
  author = {Vidal, Adrian and Wijekoon, V. B. and Viterbo, Emanuele},
  year = {2023},
  month = may,
  pages = {4452--4457},
  issn = {1938-1883},
  doi = {10.1109/ICC45041.2023.10279497},
  urldate = {2024-03-14}
}


\IEEEtriggeratref{4}

\end{document}